\documentclass[conference]{IEEEtran}
\IEEEoverridecommandlockouts
\usepackage{cite}
\usepackage{amsmath,amssymb,amsfonts}
\usepackage{graphicx}
\usepackage{textcomp}
\usepackage{xcolor}

\usepackage{url}
\usepackage{xspace}
\usepackage{multirow}
\usepackage{tikz}
\usepackage{pgfplots}
\pgfplotsset{compat=1.16}
\usepackage{subcaption}
\usepackage{algorithm}
\usepackage{algpseudocode}
\newtheorem{theorem}{Theorem}

\usepackage{booktabs}
\usepackage{colortbl}
\definecolor{bandA}{gray}{0.95}
\definecolor{bandB}{gray}{0.88}

\newcommand{\noopsort}[1]{}
\newcommand{\torchgt}{\textsc{TorchGT}\xspace}
\newcommand{\nodeformer}{\textsc{NodeFormer}\xspace}
\newcommand{\ours}{\textsc{BudgetGT}\xspace}

\def\BibTeX{{\rm B\kern-.05em{\sc i\kern-.025em b}\kern-.08em
    T\kern-.1667em\lower.7ex\hbox{E}\kern-.125emX}}
\begin{document}

\title{BudgetGT: Distributed Full-Graph Transformer on Large Graphs with Budgeted Attention and Memory}

\author{\IEEEauthorblockN{1\textsuperscript{st} Chu Zhao}
\IEEEauthorblockA{\textit{School of Computer Science and Engineering} \\
\textit{Northeastern University}\\
Shenyang, China \\
zhaoc5@mails.neu.edu.cn}
\and
\IEEEauthorblockN{2\textsuperscript{nd} Yuan Fang}
\IEEEauthorblockA{\textit{School of Computing and Information Systems} \\
\textit{Singapore Management University}\\
Singapore, Singapore \\
yfang@smu.edu.sg}
\and
\IEEEauthorblockN{3\textsuperscript{rd} Yuhai Zhao}
\IEEEauthorblockA{\textit{School of Computer Science and Engineering} \\
\textit{Northeastern University}\\
Shenyang, China \\
zhaoyuhai@mail.neu.edu.cn}
\and
\IEEEauthorblockN{4\textsuperscript{th} Zhengkui Wang}
\IEEEauthorblockA{\textit{Information and Communications Technology Cluster} \\
\textit{Singapore Institute of Technology}\\
Singapore, Singapore \\
zhengkui.wang@singaporetech.edu.sg}
}

\maketitle

\begin{abstract}
Graph Transformers have attracted increasing attention for their ability to capture nonlocal dependencies.
However, pairwise attention makes training on large graphs costly.
Although sparse attention reduces this burden, computational and memory demands can remain prohibitively high on a single GPU.
An existing distributed system extends training capacity, but its sparse execution depends on graph connectivity conditions, which real-world large graphs often fail to satisfy.
We present \textsc{BudgetGT}, a distributed full-graph training system that reduces attention cost and memory pressure without imposing graph connectivity conditions.
To bound attention computation, we combine graph neighbors with random-walk candidates under a per-node budget, retaining local and multi-hop information without full neighborhood enumeration.
A probe-guided controller adapts the budget split during training, eliminating the need for costly exhaustive searches on each dataset.
To keep construction buffers and activations within GPU memory, we model phase-specific memory peaks and construction--computation overlap.
A cost-based planner uses these estimates to select efficient edge-construction and activation-retention strategies for each workload and memory budget.
On five graphs, \textsc{BudgetGT} achieves comparable or up to $30.4$ percentage points higher accuracy than the state-of-the-art system, and trains $2.2$--$5.6\times$ faster per epoch with up to an order of magnitude less GPU memory.
Further experiments demonstrate the system's strong generality across backbones and effective memory scaling with additional GPUs, and confirm the controller's high allocation accuracy and the planner's effectiveness under diverse configurations.
\end{abstract}


\section{Introduction}
Graph Transformers~\cite{graphormer,SAN,graphgps,exphormer} let each node attend beyond its immediate neighborhood, capturing dependencies that message-passing Graph Neural Networks (GNNs) fail to propagate~\cite{alon2020bottleneck}, and reach strong accuracy on many graph learning tasks~\cite{wu2022nodeformer,wu2023sgformer,ma2023graph}.
However, dense attention considers every pair of nodes, making its computation and memory demands prohibitive on large graphs.

Existing scalable Graph Transformers reduce attention cost through kernelized or linear operators~\cite{wu2022nodeformer,choromanski2020rethinking,wu2023difformer}, compressed representations~\cite{chen2022nagphormer}, or sparse structural scopes~\cite{gt,exphormer}, but they are designed for a single GPU.
Keeping all node states and attention data on a single GPU can exhaust device memory on large graphs, forcing methods such as SGFormer~\cite{wu2023sgformer} to adopt mini-batch training.
Prior systems studies further identify growing activation memory and irregular sparse data access as bottlenecks in large-graph training~\cite{torchgt}.
Extending these methods to multiple GPUs requires additional management of cross-device dependencies, communication, and data placement~\cite{lin2026scalable}.

Distributed GNN systems~\cite{wang2022neutronstar,wang2023hongtu,song2022ecgraph,zhang2024sylvie} scale training through neighbor sampling~\cite{zheng2020distdgl}, feature caching~\cite{lin2020pagraph}, and communication optimization~\cite{gandhi2021p3,wan2022pipegcn}.
However, their graph partitioning and communication strategies are designed around native edges and require adaptation to handle the additional connections introduced by graph attention.
The state-of-the-art distributed Graph Transformer system, \torchgt~\cite{torchgt}, interleaves topology-based sparse attention with dense attention to reduce computation while
preserving model quality.
It further uses cluster-aware parallelism and a cluster-sparse layout to reduce communication and improve memory efficiency.

However, \torchgt requires connectivity conditions to keep attention sparse.
On \emph{ogbn-products}~\cite{ogb}, 97.42\% of nodes lie in a single connected component of the graph.
However, the subgraphs induced by 256K-node sequences contain a median of 63,583 components
(Table~\ref{tab:torchgt_conditions}).
\begin{table}[t]
\centering
\caption{Sequence partitioning on \emph{ogbn-products}.}
\label{tab:torchgt_conditions}
\footnotesize
\begin{tabular}{@{}lrrrr@{}}
\toprule
Sequence length & 16K & 64K & 256K & Whole graph \\
\midrule
Components & 13,544 & 35,044 & 63,583 & 52,658 \\
Excluded edges (\%) & 99.67 & 98.68 & 94.68 & 0 \\
\bottomrule
\end{tabular}
\end{table}
These sequences therefore require dense attention and exclude an average of 94.68\% of their incident graph edges because the other endpoint lies outside the sequence.
Dense attention within a sequence cannot restore these missing connections.
This raises two questions: how to keep attention cost low while retaining useful graph structure, and how can it run efficiently within GPU memory?

For the first question, we observe that trained Graph Transformers favor one-hop neighbors, but their attention to more distant nodes varies across graphs and models (Fig.~\ref{fig:hop_relative}).
For example, on \emph{ogbn-arxiv}~\cite{ogb}, Exphormer's attention per node drops sharply beyond one hop, whereas NodeFormer's shows a more gradual overall decline with distance.
These observations motivate preserving local connections while adapting the allocation to multi-hop information.
However, one-hop attention alone limits direct access to multi-hop nodes, while enumerating $k$-hop neighborhoods can cause neighborhood explosion.
We therefore propose \ours, which combines observed graph neighbors with Random-Walk (RW) candidates~\cite{node2vec,appnp} under an \emph{attention budget} of at most $B$ keys per query.
Graph neighbors preserve direct local connections, while random walks provide topology-guided multi-hop candidates without full neighborhood enumeration.
A lightweight \emph{probe-guided controller} allocates the budget between the two sources online and freezes once the split stabilizes.
As the best split varies across graphs, the controller finds a near-best split for each dataset without a separate exhaustive search.

The second question remains even with a fixed budget, as execution strategies differ in memory and time costs.
Along the memory axis, the same fixed split can have different CPU and GPU memory peaks depending on edge-construction placement and retention of intermediates.
Along the time axis, execution strategies affect both construction speed and its overlap with computation, determining how much construction time remains exposed during training.
The device's available GPU memory therefore imposes a second constraint, the \emph{memory budget}, that governs how the stabilized logical plan is physically executed.
\ours compiles the logical plan into a physical execution plan with a \emph{cost-based planner}.
Under the memory budget, the planner places each edge-construction task on the device that fits the workload, overlaps CPU construction with GPU computation when applicable, and uses activation recomputation~\cite{chen2016checkpoint} as an additional elastic reserve.
Together, these choices balance memory usage and training time, enabling efficient execution within available memory.
We separate attention-scope selection from execution planning, following the distinction between logical and physical plans in query optimization~\cite{selinger1979access}.
The controller adjusts the logical plan online, following the idea of adaptive query processing~\cite{kabra1998efficient,avnur2000eddies}, while the planner selects a physical plan through cost-based optimization over the CPU--GPU memory hierarchy.
We summarize the contributions as follows.
\begin{itemize}
    \item We develop a budgeted attention scope for distributed full-graph training, limiting each node's attention to graph neighbors and random-walk candidates under a fixed budget. 
    A probe-guided controller adapts the budget split during training, eliminating the need for costly exhaustive searches on each dataset.

    \item We show that the same attention budget can incur different construction costs and memory peaks, and design a cost-based planner that coordinates edge-construction placement and activation retention using phase-aware memory and overlap-aware time models.

    \item We evaluate \ours on five graphs and four backbones.
    Against the state-of-the-art distributed system, it achieves comparable or up to $30.4$ percentage points higher accuracy, $2.2$--$5.6\times$ faster epochs, and up to an order of magnitude less GPU memory.
\end{itemize}

\section{Background and Related Work}
\subsection{Full-Graph Training of Graph Transformers}
Let $G=(V,E)$ have $n=|V|$ nodes with features $X\in\mathbb{R}^{n\times d_{\mathrm{in}}}$.
Full-graph training processes all nodes in every forward pass rather than sampled mini-batches.
Each node $v_i$ keeps a hidden state $h_i\in\mathbb{R}^{d}$, which stacked Transformer layers update, each consisting of a multi-head attention (MHA) sublayer and a feed-forward network (FFN).
For one head with hidden states $H\in\mathbb{R}^{n\times d}$ and $Q=HW_Q$, $K=HW_K$, $V=HW_V$, dense attention is computed as follows:
\begin{equation}
\mathrm{Attn}(Q,K,V)=\mathrm{Softmax}\!\left(\tfrac{QK^\top}{\sqrt{d_k}}\right)V.
\end{equation}
Graph Transformers inject structure through node-wise encodings or a pair-wise bias in the attention logit,
\begin{equation}
A_{ij}=\tfrac{(h_iW_Q)(h_jW_K)^\top}{\sqrt{d_k}}+\phi(v_i,v_j),
\end{equation}
where $\phi$ may encode adjacency, shortest-path distance, or edge type~\cite{graphormer,SAN}.
FFN computation costs $O(nd^2)$, whereas dense pairwise attention costs $O(n^2d)$ and can become the bottleneck on large graphs.
Dense full-graph attention is prohibitive on large graphs, which motivates sparse scopes that bound memory and construction cost.

\subsection{Scalable Graph Transformers}
Graph Transformers adapt the Transformer~\cite{attention} to graphs by injecting structure into node representations~\cite{SAN}, attention scores~\cite{graphormer,ma2023graph}, or the layer composition~\cite{gt,graphgps}.
For example, Graphormer adds shortest-path and edge biases to the logits, GraphGPS pairs local message passing with global attention, and Exphormer~\cite{exphormer} restricts attention to graph and expander edges.
All-pair models such as Graphormer and SAN still scale quadratically with the number of nodes, whereas Exphormer's sparse attention is linear in the graph size.

Scalable Graph Transformers therefore reduce either the attention operator cost or the attention scope size. 
One line of work approximates all-pair attention through kernelization~\cite{choromanski2020rethinking} or compressed global representations \cite{chen2022nagphormer,kong2023goat}.
NodeFormer~\cite{wu2022nodeformer} and SGFormer~\cite{wu2023sgformer} adopt simplified or kernelized operators for scalable all-pair message passing, DIFFormer~\cite{wu2023difformer} derives a linear-complexity attention from energy-constrained diffusion, and DualFormer~\cite{dualformer} linearizes attention by reordering its matrix multiplications to make large-graph training tractable.
Another line restricts each query to a sparse structural neighborhood, as GT~\cite{gt} attends only over graph neighbors and Exphormer~\cite{exphormer} over graph and expander edges.

\subsection{Distributed Graph Learning}
Distributed graph learning has been studied extensively for message-passing GNNs~\cite{kipf2017gcn,hamilton2017graphsage}, where computation follows native graph edges or sampled neighborhoods.
Many of these systems come from the data management community, which treats training as a problem of data placement and movement.
\emph{Mini-batch} systems such as DistDGL~\cite{zheng2020distdgl} and PaGraph~\cite{lin2020pagraph} scale training through neighbor sampling and feature caching, ByteGNN~\cite{zheng2022bytegnn} cuts the sampling cost, P3~\cite{gandhi2021p3} pipelines computation with communication, and XGNN~\cite{tang2024xgnn} and Hyperion~\cite{sun2025hyperion} extend storage to a global memory store or SSDs.
\emph{Full-graph} systems instead process the entire graph in every step and parallelize it across devices.
ROC~\cite{jia2020improving} schedules partitioned propagation across GPUs, NeutronStar~\cite{wang2022neutronstar} chooses between caching and communicating vertex dependencies, and HongTu~\cite{wang2023hongtu} keeps vertex data in host memory to train beyond GPU memory.
PipeGCN~\cite{wan2022pipegcn}, BNS-GCN~\cite{wan2022bns}, and Sancus~\cite{peng2022sancus} cut inter-partition communication by staleness or sampling, and EC-Graph~\cite{song2022ecgraph} and SNI-GNN~\cite{yu2026snignn} do so by message compression or in-network embedding prediction.

Graph Transformers attend over a much broader scope, from all-pair attention~\cite{graphormer,SAN} to sparse or linear variants~\cite{gt,exphormer,choromanski2020rethinking}, so training them at scale must also manage attention-scope construction and its temporary edge-building workspaces.
\torchgt~\cite{torchgt}, the state-of-the-art distributed system, stays efficient only while connectivity conditions keep its attention sparse, and otherwise falls back to a cost quadratic in the sequence length.
A more recent parallel training system~\cite{lin2026scalable} adapts communication schemes for sparse attention, but attends only over the native topology, leaving the scope fixed.
GNN systems manage data access along native graph edges, while existing Graph Transformer systems primarily optimize the execution of predefined attention patterns.
\ours instead decides which attention edges to materialize and how to execute the resulting workload through logical and physical planning, inspired by cost-based query optimization~\cite{selinger1979access} and adaptive query processing~\cite{kabra1998efficient,avnur2000eddies}.
Cost-based optimization has also been applied to machine learning programs, e.g., SystemML~\cite{boehm2014systemml} chooses execution plans under memory budgets.
It optimizes linear-algebra operators, whereas \ours plans the construction and placement of attention edges and the retention of activations.

\section{Adaptive Logical Planning under the Attention Budget}
\label{sec:mixed_budget_scope}
This section addresses the \emph{logical} question of which sparse edges each query should attend to. 
Motivated by the attention locality observed in trained Graph Transformers, we construct a query-centric scope using edges from two sources, real graph neighbors and random-walk candidates.
Each query has an \emph{attention budget} $B$, which limits the number of nodes it may attend to. 
During training, we adaptively allocate this budget between the two sources, much as an adaptive query processor re-optimizes a running plan~\cite{kabra1998efficient,avnur2000eddies}.

\subsection{Empirical Motivation: Attention Locality}
\label{sec:motivation}

To guide the construction of a bounded attention scope, we analyze how each query distributes attention across keys to identify patterns in trained Graph Transformers.
\begin{figure}[t]
\centering
\includegraphics[width=\linewidth]{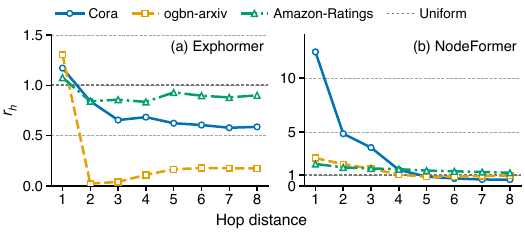}
\caption{Hop-relative attention $r_h$ in the last layer after training, for (a) Exphormer and (b) \nodeformer. The dotted line $r_h=1$ is uniform attention.}
\label{fig:hop_relative}
\end{figure}

\paragraph{Learned locality across graphs and models}
We examine two models that train on the full graph and let each query attend beyond its neighbors.
Exphormer~\cite{exphormer} computes sparse attention over graph edges and random expander edges, and \nodeformer~\cite{wu2022nodeformer} approximates all-pair attention with a kernel.
The hop-relative attention $r_h$ is the average attention that a query assigns to a key $h$ hops away, divided by its value under uniform attention, so $r_h=1$ means no preference.
On \emph{Cora}~\cite{planetoid}, \emph{ogbn-arxiv}~\cite{ogb}, and the heterophilic \emph{Amazon-Ratings}~\cite{platonov2023critical}, both models assign one-hop neighbors more than uniform attention (Fig.~\ref{fig:hop_relative}).
In Exphormer, $r_h$ drops below $1$ at every hop beyond the first, and in \nodeformer it decays gradually with distance.
The bias is learned, as $r_h\approx1$ at every hop at initialization and nearby hops are up-weighted as training proceeds.

\paragraph{Graph-dependent long-range attention}
Attention to distant nodes varies across graphs and models.
Beyond the first hop, Exphormer's $r_h$ stays near $0.9$ on \emph{Amazon-Ratings} and falls to about $0.6$ on \emph{Cora}, but drops below $0.2$ on \emph{ogbn-arxiv}.
For \nodeformer, $r_h$ stays above $1$ at every hop on \emph{Amazon-Ratings}, while on \emph{ogbn-arxiv} it dips at hop~5 and returns to $1$ by hop~8.
The attention assigned to distant nodes thus depends on the graph and the model, so topology alone cannot determine it in advance.

\paragraph{Implications for the scope}
A scope should therefore keep each query's one-hop neighbors, where attention concentrates, and add multi-hop candidates for long-range attention, whose share cannot be fixed in advance.
Two questions follow: how to generate multi-hop candidates whose number stays bounded, and how to divide each query's budget between the two kinds during training.

\subsection{From Local Edges to Random-Walk Edges}
\label{sec:rw_motivation}
\begin{figure}[htbp]
    \centering
    \begin{subfigure}{0.47\linewidth}
        \centering
        \includegraphics[width=\linewidth]{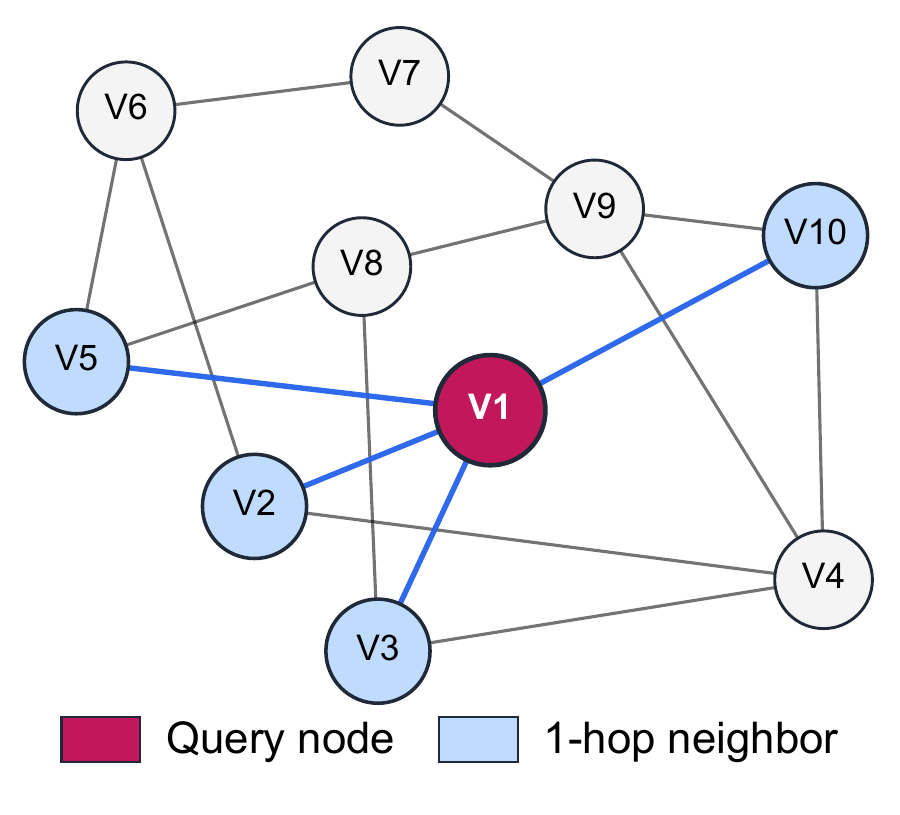}
        \caption{Observed real edges}
        \label{fig:real_edges}
    \end{subfigure}
    \hfill
    \begin{subfigure}{0.47\linewidth}
        \centering
        \includegraphics[width=\linewidth]{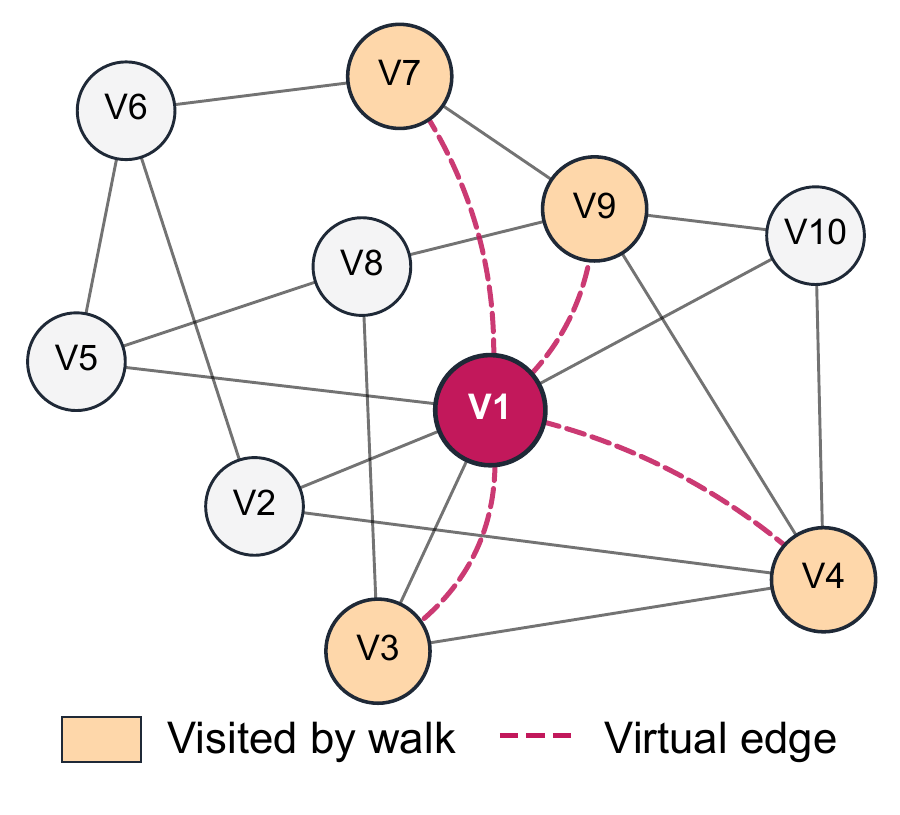}
        \caption{Generated virtual edges}
        \label{fig:rw_virtual_edges}
    \end{subfigure}
    \caption{Query-centric attention scope construction.
    The red node denotes the query node, blue nodes denote observed one-hop neighbors, and dashed edges denote virtual attention edges generated from random-walk paths.}
    \label{fig:mixed_scope}
\end{figure}
A fixed $k$-hop neighborhood would admit multi-hop candidates, but its size can grow exponentially in $k$, which gives uneven scopes and unstable memory in full-graph training, where the candidate edges of every node must be generated during training.
Random walks, long used to learn graph representations~\cite{node2vec} and to model multi-hop reachability~\cite{appnp}, lie between one-hop and $k$-hop scopes.
They follow the topology to multi-hop nodes without enumerating the $k$-hop neighborhood, and the number and length of walks bound the number of candidates.
Each query's scope combines its observed real edges with \emph{virtual} edges generated from its walks (Fig.~\ref{fig:mixed_scope}).
Virtual edges are used only for attention and leave the topology unchanged.

\paragraph{Two candidate pools}
Following this design, we build two candidate pools for each query node $q$.
The real pool $\mathcal N_q$ holds its observed one-hop neighbors and is built once.
The walk pool $\mathcal W_q$ holds the visits of $R$ random walks of length $L$ that start at $q$, and it is redrawn every epoch with a per-epoch seed.
$\mathcal W_q$ is a list of at most $RL$ entries rather than a set, so a node that the walks visit several times appears several times.

\paragraph{Nested selection}
A split must be cheap to apply and easy to compare with its neighbors.
Given a split $\mathbf b=(b_r,b_w)$ with $b_r{+}b_w{=}B$ and a sampling seed $\xi$, each query ranks each pool by a seeded hash and keeps the first $b_r$ real and the first $b_w$ walk entries, or the whole pool if it is shorter.
The rule is local to each query, so it needs no global coordination, and it is \emph{nested}.
Under the same seed, raising $b_r$ or $b_w$ by a step of $\Delta$ slots only appends the next $\Delta$ entries of that pool, so adjacent splits share all other candidates.
This is what lets Section~\ref{sec:budget_allocation} compare them with little sampling noise.

\paragraph{Set-valued scope}
With $\mathcal S^{\mathrm{real}}_q(\mathbf b)$ and $\mathcal S^{\mathrm{rw}}_q(\mathbf b)$ the retained real and walk entries, attention reads the distinct retained keys,
\begin{equation}
 \mathcal S_q(\mathbf b)=\mathcal S^{\mathrm{real}}_q(\mathbf b)\cup
 \operatorname{set}\bigl(\mathcal S^{\mathrm{rw}}_q(\mathbf b)\bigr),\qquad
 |\mathcal S_q(\mathbf b)|\le B,
 \label{eq:merged_scope}
\end{equation}
with an ordinary softmax over $\mathcal S_q(\mathbf b)$.
Every split thus uses the same attention operator, and the workload is bounded by $nB$ edges for $n$ query nodes.
The bound is not always reached.
A repeated walk visit or a key retained by both pools adds no new key, and a short pool cannot fill its share, so two splits with the same $B$ can expose different numbers of keys and incur different execution costs.

\paragraph{Attention complexity}
Each attention layer over these scopes takes $O(nBd)$ time, against $O(n^2 d)$ for dense attention and $O(|E|d)$ for attention over all graph edges, and stores $O(nB)$ attention scores per head instead of $O(|E|)$.
Its cost therefore follows the budget rather than the graph's density.
Sequence parallelism (SP) further divides both costs across devices.

\subsection{Probe-Guided Budget Allocation}
\label{sec:budget_allocation}

The best split depends on the graph and the learned attention.
On \emph{ogbn-arxiv}, Graphormer does best with real edges only, whereas on \emph{genius} mixed splits do best.
Training one model per split is costly, but a split changes only which keys attention reads, not the model's parameters, so one model can be evaluated under every split and the split chosen during training.
A probe-guided controller does so as a local search.
Between training epochs, it runs \emph{probe rounds} that compare the current split $\mathbf b_t$ with its neighbors, the splits one $\Delta$ step away.

\paragraph{Paired comparison}
Adjacent splits often differ by less than the sampling noise, so the controller compares them on the same random draw.
In round $t$, it draws one sampling seed $\xi_t$ and evaluates the current model $\theta_t$ under every candidate $a$ on a held-out probe set $V_p$ (the validation split), with no gradient step in between.
With $\ell(\theta_t,a;\xi_t)$ the probe loss, the score of a candidate is its relative gain over the current split,
\begin{equation}
 \widehat g_t(a)=\frac{\ell(\theta_t,\mathbf b_t;\xi_t)-\ell(\theta_t,a;\xi_t)}
                      {\ell(\theta_t,\mathbf b_t;\xi_t)},
 \label{eq:paired_gain}
\end{equation}
with expectation $g_t(a)$ over $\xi$.
Here, $g_t(a)>0$ means that, under the current model, $a$ has a positive expected relative gain over the current split.
Sharing $\xi_t$ is the common-random-numbers technique~\cite{glasserman1992crn}, which reduces the variance of the difference from $\operatorname{Var}\ell_{\mathbf b}+\operatorname{Var}\ell_a$ to $\operatorname{Var}\ell_{\mathbf b}+\operatorname{Var}\ell_a-2\operatorname{Cov}(\ell_{\mathbf b},\ell_a)$.
By nestedness, the two scopes differ in at most $2\Delta$ of up to $B$ entries per query, so the covariance is large and the difference mainly reflects the moved block.
Probe edges are built only for queries within the model's receptive field of $V_p$.
They also drop walk entries that point to a real neighbor, so a walk move is judged only by the keys that the real pool cannot supply.
Training keeps these entries because the real pool retains only $b_r$ neighbors, so a walk that reaches any other neighbor still adds a key.

\paragraph{Choosing a move}
The rule depends on whether the budget is full.
While it is not (Phase~I), every move adds $\Delta$ slots, so even a wrong choice adds capacity, and the probe only picks which source grows.
It grows the source with the larger gain if that gain exceeds an indifference threshold $\tau$. Otherwise it repeats the last such choice, or keeps the split balanced if there is none.
Growth predicted to exceed device memory is never proposed.
Once the budget is full (Phase~II), a move shifts $\Delta$ slots from one source to the other.
Loss ranks such moves while they differ clearly, but near the best split the differences fall below the noise, so accuracy serves as a check.
A candidate wins a round if its gain is the largest and exceeds $\tau$, unless its probe accuracy is more than two binomial standard errors below the current split's on $V_p$.
If every gain lies within $\pm\tau$, a candidate wins only with a significantly higher accuracy, and otherwise the split stays.
Later moves have smaller gains, closer to the noise, so each commit needs one more consecutive win than the previous one, up to $K_{\max}$.

\paragraph{When to stop}
The controller freezes the split once probes stop finding a better one, i.e., after $P$ rounds without a winner since the last commit.
Rounds before the first gain above $\tau$ do not count, because an untrained model cannot yet tell splits apart.
Section~\ref{sec:allocation_analysis} bounds the probability of a wrong commit and of freezing too early.

\begin{algorithm}[t]
\caption{Probe-Guided Budget Allocation}
\label{alg:budget_allocation}
\begin{algorithmic}[1]
\Require budget $B$, step $\Delta$, probe set $V_p$, threshold $\tau$,
  evidence cap $K_{\max}$, patience $P$
\Ensure trained model $f_\theta$ and frozen split $\mathbf b^*$
\State $\mathbf b\gets(\Delta,\Delta)$;\ $j,c,u\gets0$;\ $d_{\mathrm{last}}\gets\bot$
\For{epoch $t=1,2,\ldots$}
  \State train one epoch on the scopes $\mathcal S_q(\mathbf b)$
  \If{the model is warming up or $\mathbf b$ is frozen}
    \State \textbf{continue}
  \EndIf
  \State draw $\xi_t$; compute $\widehat g_t(a)$ on $V_p$ for each neighbor $a$ of $\mathbf b$
  \If{$b_r+b_w<B$}
    \State $d\gets$ growth direction \Comment{Phase I}
  \Else
    \State $d\gets$ round winner, or $\bot$ if none \Comment{Phase II}
  \EndIf
  \If{$d=\bot$}
    \State $c\gets0$;\ $u\gets u+1$ once a gain has exceeded $\tau$;\ freeze $\mathbf b$ if $u\ge P$
  \Else
    \State $c\gets c+1$ if $d=d_{\mathrm{last}}$, else $c\gets1$;\ $d_{\mathrm{last}}\gets d$
    \If{$c\ge\min(j+1,K_{\max})$}
      \State apply $d$ to $\mathbf b$;\ $j\gets j+1$;\ $c,u\gets0$;\ $d_{\mathrm{last}}\gets\bot$
    \EndIf
  \EndIf
\EndFor
\State \Return $f_\theta$ and $\mathbf b^*\gets\mathbf b$
\end{algorithmic}
\end{algorithm}

Algorithm~\ref{alg:budget_allocation} proceeds in three stages.
\emph{Initialization} (line~1) starts from $(\Delta,\Delta)$ with zero counters, where $j$ counts committed moves, $c$ consecutive wins of the last winner $d_{\mathrm{last}}$, and $u$ rounds without a winner since the last commit.
\emph{Probing} (lines~4--12) scores the neighbors after each epoch, except during warm-up and after freezing, and proposes the growth direction in Phase~I or the round winner in Phase~II.
\emph{Confirmation} (lines~13--20) commits a move after $\min(j+1,K_{\max})$ consecutive wins and freezes the split after $P$ rounds without a winner.
We set $\Delta=2$, $\tau=3\times10^{-5}$, $K_{\max}=3$, and $P=3$ as our default settings.

\subsection{Analysis of the Controller}
\label{sec:allocation_analysis}

Since the controller acts on noisy gains, it may commit a move that does not help, miss one that clearly helps, or freeze too early.
We bound all three in Phase~II and exclude Phase~I, as its moves only add slots.
\textbf{(A1)}~Given the history, the probe error $\widehat g_t(a)-g_t(a)$ is $\sigma$-sub-Gaussian (tails no heavier than a Gaussian's of standard deviation $\sigma$), and errors in different rounds are independent, as each round draws a fresh $\xi_t$.
\textbf{(A2)}~Within the rounds considered, each candidate's true gain keeps its sign, i.e., the model changes more slowly than the controller confirms.
A candidate $a$ is \emph{non-improving} if $g_t(a)\le0$, and \emph{$\delta$-better} if $g_t(a)$ exceeds both $\tau$ and every other candidate's gain by at least $\delta$.
A round has $m\le2$ candidates besides the current split, and $\epsilon$ bounds the probability that the accuracy rule elects a non-improving candidate or vetoes one with gain above $\tau$.
Per round, $\alpha$ below bounds the chance that a non-improving candidate wins, and $\beta$ that a $\delta$-better one loses:
\begin{equation}
 \alpha=e^{-\tau^2/2\sigma^2}+\epsilon,\qquad
 \beta=m\,e^{-\delta^2/8\sigma^2}+\epsilon .
 \label{eq:pq}
\end{equation}

\begin{theorem}[Sequential confirmation]
\label{thm:sequential_confirmation}
Under (A1) and (A2), in Phase~II:
\begin{enumerate}
\item \emph{False commits.} A candidate that is non-improving throughout $K$ consecutive rounds wins all of them with probability at most $\alpha^K$.
\item \emph{Detection.} A candidate that is $\delta$-better throughout $K$ consecutive rounds wins all of them, and is committed if $K$ confirmations are required, with probability at least $(1-\beta)^K$.
\item \emph{Premature freezing.} If some candidate is $\delta$-better in each of the $W$ rounds after a commit, the controller freezes within them with probability at most $\binom{W}{P}\beta^{P}$.
\end{enumerate}
\end{theorem}

\begin{IEEEproof}
By (A1) and (A2), per-round bounds multiply over consecutive rounds (condition on the history before the last round and induct).
(1)~A non-improving $a$ ($g_t(a)\le0$) wins on loss only if its error exceeds $\tau$, with probability at most $e^{-\tau^2/2\sigma^2}$, and on accuracy with probability at most $\epsilon$, so a round has probability at most $\alpha$.
(2)~A $\delta$-better $a$ loses a round in only three ways: its own error falls below $-\delta$, a competitor's error exceeds that of $a$ by more than $\delta$, or $a$ is vetoed.
The difference of two $\sigma$-sub-Gaussian errors is $2\sigma$-sub-Gaussian, so each of the first $m$ ways has probability at most $e^{-\delta^2/8\sigma^2}$, and the veto at most $\epsilon$, so a round fails with probability at most $\beta$.
(3)~Freezing needs $P$ rounds without a winner after the last commit, and while a $\delta$-better candidate exists, each such round is a failure as in (2).
A union bound over the $\binom{W}{P}$ ways to place these rounds among the $W$ completes the proof.
\end{IEEEproof}

\paragraph{What the analysis implies}
Each additional confirmation multiplies the false-commit probability bound by $\alpha$.
Later moves need more consecutive wins because their gains are harder to distinguish from noise, while early moves with large gains need only one.
Waiting $P$ rounds reduces premature freezing, while using the same random seed reduces noise $\sigma$ and tightens all three bounds.

\paragraph{Selection cost}
Each probe round adds at most two forward passes over the receptive field of $V_p$ and no gradient step, and its cost does not grow with the number of splits.
Exhaustive search would instead train one model per split, $B/\Delta+1$ models in total.

\section{Cost-Based Physical Planning under the Memory Budget}
\label{sec:physical_planning}

Section~\ref{sec:mixed_budget_scope} produces the logical plan $\omega^*=(b_r^*,b_w^*,L,R)$ under the attention budget, which fixes the budgeted mix of real and random-walk edges exposed to attention.
It does not say \emph{how} that scope should be constructed, materialized, and executed.
Without careful physical planning, the memory footprint of full-graph training can exceed device capacity and trigger out-of-memory failures.

This section introduces the second governing constraint, the \emph{memory budget}, i.e., the available GPU High Bandwidth Memory (HBM) and host memory, $C_{\mathrm{HBM}}$ and $C_{\mathrm{host}}$.
\ours casts physical planning as cost-based plan selection~\cite{selinger1979access} under the memory budget, as SystemML does for machine learning programs~\cite{boehm2014systemml}.
The planner first expands the set of configurations that are trainable under the memory budget, then selects the fastest feasible one.

\subsection{Same Logical Plan, Different Physical Costs}
\label{sec:workload_heterogeneity}

The total budget $B$ bounds the nominal attention degree, but it is not a physical cost model. 
Real-edge selection, random-walk generation, and edge merging/deduplication use different operators and act on different tensors. 
As a result, splits with the same $B$ can have very different construction times and peak memory.
  
Table~\ref{tab:workload_char} shows this mismatch. 
All measurements use full activation recomputation so that the reported memory peaks reflect construction cost rather than cached activations.
Building on the GPU is uniformly fast (construction stays under $1$\,s), but its temporary workspace inflates the memory peak to about $13$\,GiB, which competes directly with model states and moves the construction-memory pressure onto HBM. 
However, CPU construction stores its buffers in host memory, keeping peak GPU memory usage at the model states.
CPU construction is much slower, especially when real edges are included.
On \emph{Amazon}, construction time rises from $1.5$ to $11.7$\,s because sorting the full edge set dominates the cost.

\begin{table}[htbp]
\centering
\caption{Execution strategy of $4$-layer Graphormer$_{\mathrm{Slim}}$
($B{=}8$, $\mathrm{SP}{=}4$).
\emph{Prep} is the CPU/GPU edge-construction time (s).
Each \emph{Memory} entry is the (CPU RAM, GPU HBM) peak in GiB under that build
device.
\emph{Comp.} is the per-step model computation time (s), independent of the build device and split, that CPU preparation can overlap with.}
\label{tab:workload_char}
\begin{tabular}{ll c cc c}
\toprule
\multirow{2}{*}{Dataset} & \multirow{2}{*}{$(b_r,b_w)$}
  & \multirow{2}{*}{\shortstack{Prep\\\scriptsize CPU/GPU}}
  & \multicolumn{2}{c}{Memory (RAM, HBM)} & \multirow{2}{*}{\shortstack{Comp.\\\scriptsize (s)}} \\
\cmidrule(lr){4-5}
 & & & CPU build & GPU build & \\
\midrule
\multirow{3}{*}{Amazon}
 & $(0,8)$ &  1.5/0.2 & (20.2,\,5.7)  & (4.8,\,13.6) & \multirow{3}{*}{6.6} \\
 & $(4,4)$ & 11.7/0.6 & (20.2,\,5.5)  & (4.8,\,13.6) & \\
 & $(8,0)$ & 10.7/0.8 & (20.2,\,5.2)  & (4.8,\,13.6) & \\
\midrule
\multirow{3}{*}{products}
 & $(0,8)$ &  3.6/0.3 & (11.4,\,10.6) & (2.7,\,13.6) & \multirow{3}{*}{10.4} \\
 & $(4,4)$ &  8.0/0.5 & (11.5,\,10.3) & (2.7,\,13.4) & \\
 & $(8,0)$ &  5.9/0.4 & (11.5,\,10.0) & (2.7,\,12.8) & \\
\bottomrule
\end{tabular}
\end{table}
Raw construction time does not necessarily add to training time. 
The CPU build runs on a separate resource and can overlap with the previous step's GPU \emph{Compute}, so only the part that exceeds the compute window is exposed. 
On \emph{ogbn-products}, CPU preparation ($4$--$8$\,s) overlaps almost entirely with the $\sim$$10$\,s compute window, adding little transfer delay while keeping peak GPU memory usage low.
On \emph{Amazon}'s real-heavy splits, however, CPU preparation ($\sim$$11$\,s) exceeds the $\sim$$7$\,s window and cannot be hidden, so GPU construction wins despite its memory cost. 
Neither location is best everywhere as the choice depends on construction speed, memory headroom, and the overlap window.

The overlap window itself depends on the model, not the workload.
For example, doubling the depth from $4$ to $8$ layers widens \emph{Amazon}'s compute window from $\sim$$7$\,s to $\sim$$13$\,s while leaving CPU construction unchanged, so CPU construction becomes the better choice.
More broadly, the best strategy shifts with the dataset, the budget split, the model depth, the sequence-parallel degree, and the host core count.
We therefore design a cost-based planner that selects an execution plan based on the actual workload and available resources.

\subsection{Phase-Aware Memory Coordination}
\label{sec:memory_coordination}

Full-graph training holds memory objects with different lifetimes. Parameters, optimizer states, and optionally cached source topology persist across epochs. 
The materialized attention edges stay live during the model step. Edge construction also creates short-lived sorting, random-walk, and merge workspaces. 
Backward states occupy the remaining HBM during training. 
Treating these objects as a single additive footprint is overly conservative, since construction and model execution happen in different GPU phases. 
Optimizing them independently is unsafe, since the persistent objects coexist with both phases.
  
For a physical plan $p=(\pi,\rho)$, where $\pi$ is the edge policy and $\rho$ is the model-state reserve, \ours estimates the peak as
\begin{equation}
\begin{aligned}
\widehat{M}_{\mathrm{peak}}(p)
&=\max\bigl\{\, M_{\mathrm{build}}(\pi),\;
   M_{\mathrm{base}}+M_{\mathrm{topo}}(\pi) \\
&\qquad\qquad +\,M_{\mathrm{attn}}+M_{\mathrm{state}}(\rho) \,\bigr\} \\
&\le C_{\mathrm{HBM}}.
\end{aligned}
\label{eq:phase_memory}
\end{equation}
Here $M_{\mathrm{build}}$ is the measured peak of edge construction, and the second term models the training phase, where $M_{\mathrm{base}}$ is the plan-independent state, such as parameters and optimizer states, $M_{\mathrm{topo}}$ the optional persistent topology, $M_{\mathrm{attn}}$ the realized attention edges, and $M_{\mathrm{state}}$ the backward-state reserve.
A CPU-prefetch plan must also fit its host-side construction buffers within $C_{\mathrm{host}}$, which is far cheaper than HBM.

\paragraph{Edge-construction paths}
The two edge sources differ sharply in construction cost. 
Real edges come from a lightweight top-$k$ selection over observed neighbors, whereas random-walk edges require walking over a copy of the full topology and deduplicating the visits, so their transient working set scales with the graph's edge count and can dominate HBM. 
On the edge-dense \emph{Amazon} graph, for example, it reaches $12.5$\,GiB.
Where this working set is placed therefore matters most for fitting the budget, and the planner chooses among four edge paths.
\emph{GPU persistent} caches source topology on the GPU and builds all edges on device.
\emph{GPU ephemeral} also builds on device but stages source data on demand.
Both build the walks on the GPU only when their working set fits the HBM budget, and fall back to the CPU otherwise.
\emph{CPU rank-local prefetch} and \emph{CPU broadcast prefetch} build edges on the CPU, on every rank or on one rank that broadcasts them, respectively.
Both overlap construction with the previous step's GPU compute, hiding it whenever the compute window is wide enough.

\paragraph{Elastic model-state reserve}
The model exposes three backward-state levels built on activation recomputation~\cite{chen2016checkpoint}, a technique that TSPLIT~\cite{nie2022tsplit} applies at sub-tensor granularity to DNNs.
A layer is recomputed in full ($D_{\mathrm{rec}}$ layers), keeps its attention states while recomputing the FFN ($D_{\mathrm{mha}}$), or is retained in full ($D_{\mathrm{ret}}$), with $D_{\mathrm{rec}}{+}D_{\mathrm{mha}}{+}D_{\mathrm{ret}}{=}D$ for a $D$-layer model.
We treat retention as a \emph{secondary} choice, using only the HBM left by the selected edge path.
The planner recomputes more layers when edge construction is memory-heavy, and retains more states when it is not.

\subsection{Cost-Based Plan Selection}
\label{sec:cost_model}

We defer physical profiling until the controller in Section~\ref{sec:budget_allocation} freezes $\omega^*$. 
During budget exploration, \ours uses CPU broadcast construction and full recomputation as a conservative execution baseline. 
Once the logical workload stabilizes, the probe-only caches are released and the planner measures the policies on the actual real/RW split. 
This ordering keeps the system from optimizing a temporary workload that disappears after the next budget update. 

\begin{figure}[t]
\centering
\includegraphics[width=\linewidth]{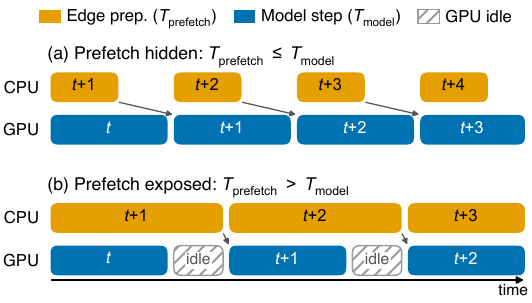}
\caption{Prefetch under two workloads. Edge preparation for epoch $t{+}1$ overlaps with GPU training of epoch $t$. \textbf{(a)} When $T_{\mathrm{prefetch}}\le T_{\mathrm{model}}$, preparation is fully hidden. \textbf{(b)} Otherwise, the uncovered preparation time creates a pipeline bubble.}
\label{fig:cpu_prefetch_overlap}
\end{figure}

\paragraph{Overlap-aware edge profiling}
CPU construction is not necessarily on the critical path. 
As shown in Fig.~\ref{fig:cpu_prefetch_overlap}, \ours constructs attention edges for the next epoch while the GPU runs the current one, much as GNN systems pipeline communication with computation~\cite{gandhi2021p3,wan2022pipegcn}. 
If preparation finishes within the model-computation window, its cost is fully hidden, and otherwise only the excess preparation time stalls the GPU. 
For each policy, the planner records the serial staging time, the overlappable preparation time, the build peak, the persistent topology bytes, and the host-memory growth.

For plan $p=(\pi,\rho)$, the exposed epoch time is estimated from the model time $T_{\mathrm{model}}(\rho)$, the serial staging time $T_{\mathrm{serial}}(\pi)$, and the overlappable preparation time $T_{\mathrm{prefetch}}(\pi)$ as
\begin{equation}
\widehat{T}_{\mathrm{step}}(p)=
T_{\mathrm{serial}}(\pi)
+\max\bigl(T_{\mathrm{model}}(\rho),\,T_{\mathrm{prefetch}}(\pi)\bigr).
\label{eq:overlap_time}
\end{equation}
For GPU policies, construction is serialized before attention and counted in $T_{\mathrm{serial}}$, so $T_{\mathrm{prefetch}}$ is zero. 
For CPU policies, the $\max$ hides preparation that fits within the model window and exposes only the excess. 
A slower CPU builder can still beat GPU construction when the model gives a large enough overlap window.

\paragraph{Lightweight state calibration and selection}
The planner tests full recomputation and a few activation-retention plans to estimate their HBM requirements.
Limiting the tested plans keeps calibration overhead low and leaves edge-path selection as our main planning decision.

Plan selection is feasibility-first. 
\ours filters policies with Eq.~\eqref{eq:phase_memory} and the host-memory limit, then ranks the surviving edge/state combinations by Eq.~\eqref{eq:overlap_time}. 
When estimated times are similar, we prefer the plan with more HBM headroom.
A committed CPU-prefetch plan is re-checked against its measured step time until two consecutive checks confirm it, so that host-load jitter does not lock in a slower plan.
Profiling runs inside normal training and, beyond building each path once, only adds timers, so its amortized overhead is negligible.

\begin{algorithm}[t]
\caption{Cost-Based Physical Plan Selection}
\label{alg:physical_planning}
\begin{algorithmic}[1]
\Require Stabilized logical plan $\omega^*$, edge policies $\mathcal{P}$
(the edge paths of Section~\ref{sec:memory_coordination}), state levels $\mathcal{R}$, host capacity $C_{\mathrm{host}}$, HBM
capacity $C_{\mathrm{HBM}}$
\Ensure Selected physical plan $p^*=(\pi^*,\rho^*)$
\State Release probe-only caches
\ForAll{$\pi\in\mathcal{P}$}
    \State Profile construction time and CPU/GPU memory peaks
\EndFor
\ForAll{$\rho\in\mathcal{R}$}
    \State Profile model time and backward-state memory
\EndFor
\State $\mathcal{F}\gets\emptyset$
\ForAll{$(\pi,\rho)\in\mathcal{P}\times\mathcal{R}$}
    \State Estimate $\widehat{M}_{\mathrm{host}}(\pi)$ and
    $\widehat{M}_{\mathrm{peak}}(\pi,\rho)$
    \If{$\widehat{M}_{\mathrm{host}}(\pi)\le C_{\mathrm{host}}$ and
    $\widehat{M}_{\mathrm{peak}}(\pi,\rho)\le C_{\mathrm{HBM}}$}
        \State Estimate $\widehat{T}_{\mathrm{step}}(\pi,\rho)$ using
        Eq.~\eqref{eq:overlap_time}
        \State $\mathcal{F}\gets\mathcal{F}\cup\{(\pi,\rho)\}$
    \EndIf
\EndFor
\State $p^*\gets\arg\min_{(\pi,\rho)\in\mathcal{F}}
\widehat{T}_{\mathrm{step}}(\pi,\rho)$
\State Synchronize $p^*$ across sequence-parallel ranks
\State \Return $p^*$
\end{algorithmic}
\end{algorithm}

Algorithm~\ref{alg:physical_planning} first frees the probe-only caches of budget exploration (line~1) and profiles each edge policy and each backward-state level separately, recording construction time, model time, and CPU/GPU memory peaks (lines~2--7).
It then keeps the combinations whose host footprint and phase-aware GPU peak (Eq.~\eqref{eq:phase_memory}) fit $C_{\mathrm{host}}$ and $C_{\mathrm{HBM}}$, estimates their epoch time with Eq.~\eqref{eq:overlap_time} (lines~8--15), and commits the fastest one on every sequence-parallel rank (lines~16--18).

\section{Experiments}

\subsection{Experimental Setup}

\paragraph{Baselines and models}
We compare against \torchgt~\cite{torchgt}, the state-of-the-art system for distributed Graph Transformer training.
Both systems run two representative models, Graphormer~\cite{graphormer} and GT~\cite{gt}, in the Graphormer$_{\mathrm{Slim}}$ and GT configurations of the \torchgt paper.
To show generality, we also implement GraphGPS~\cite{graphgps} and Exphormer~\cite{exphormer}, two scalable Graph Transformer designs.

\paragraph{Datasets}
\begin{table}[]
\caption{Dataset statistics. Tables abbreviate ogbn-arxiv, ogbn-products, and snap-patents as arxiv, products, and snap-pat.}
\label{tab:dataset}
\begin{tabular}{ccccc}
\hline
Dataset       & \# Nodes   & \# Edges    & \# Feat & \# Labels \\ \hline
ogbn-arxiv~\cite{ogb}    & 169,343   & 1,166,243   & 128          & 40        \\
ogbn-products~\cite{ogb} & 2,449,029 & 61,859,140  & 100          & 47        \\
Amazon~\cite{graphsaint}        & 1,569,960 & 264,339,468 & 200          & 107       \\ \hline
genius~\cite{linkx}        & 421,961   & 984,979     & 12           & 2         \\
snap-patents~\cite{linkx}  & 2,923,922 & 13,975,788  & 269          & 5         \\ \hline
\end{tabular}
\end{table}
Table~\ref{tab:dataset} lists the datasets we use, which span several domains, such as citation and co-purchase graphs. 
The selection includes both homophilic and heterophilic graphs. We report test accuracy on all datasets except \emph{genius}, for which we report ROC-AUC following~\cite{linkx}.
Both metrics are given in percent, with ROC-AUC multiplied by $100$, so every difference between them is in percentage points (pp).

\paragraph{Implementation and hardware}
\ours is implemented on top of PyTorch v2.1~\cite{paszke2019pytorch} with torch.distributed over NCCL~\cite{nccl}. 
All time and memory results are measured on a server with an AMD EPYC 7763 64-core CPU, 512\,GB of host memory, and four NVIDIA L40 GPUs.

\subsection{Distributed Training Comparison}
\torchgt\footnote{We build on the public release at \url{https://github.com/zxmeng98/torchgt}, adding the missing connectivity condition check.} trains on node sequences cut from the graph, at most 16K nodes long on our hardware, whereas \ours attends over the full graph.
For \ours, we set $(R,L,B)=(2,4,8)$ and cap GPU memory usage near that of \torchgt.
We match \torchgt's hyperparameters except for a higher learning rate to account for fewer updates in full-graph training.
Both systems train for 2000 epochs on four GPUs using \torchgt's data split and random seed.

\begin{table*}[htbp]
\centering
\caption{Distributed training: \torchgt vs.\ \ours on five datasets with two backbones.
  Performance is converged test accuracy (\%), or ROC-AUC (\%) on
  \emph{genius}. \emph{Diff.} is \ours minus \torchgt in pp. \emph{Full} is full-graph training on every real neighbor with no budget and no RW edges.
  OOM marks runs that exceed the 24\,GiB cap.}
\label{tab:distributed}
\begin{tabular}{cc|cccc|cccc|cccc}
\hline
\multirow{2}{*}{Dataset}   & \multirow{2}{*}{Model} & \multicolumn{2}{c}{Training (s)} & \multicolumn{2}{c}{Inference (s)} & \multicolumn{4}{c}{Performance} & \multicolumn{2}{c}{Allocated GPU (GiB)} & \multicolumn{2}{c}{Peak CPU (GiB)} \\ \cline{3-14}
                           &                        & TorchGT          & Ours          & TorchGT           & Ours          & TorchGT  & Full   & Ours   & Diff. & TorchGT             & Ours              & TorchGT           & Ours           \\ \hline
\rowcolor{bandA}                             & Graph.                 & 1.75             & 0.68          & 1.6               & 0.17          & 57.96    & 72.80  & 71.65   & +13.69   & 20.59               & 1.85              & 1.47              & 2.32           \\
\rowcolor{bandA} \multirow{-2}{*}{arxiv}     & GT                     & 1.74             & 0.35          & 1.51              & 0.24          & 55.22    & 72.41  & 71.97   & +16.75   & 20.62               & 2.46              & 1.46              & 2.05           \\
\rowcolor{bandB}                             & Graph.                 & 24.99            & 10.41         & 109.6             & 1.62          & 77.45    & OOM    & 88.20  & +10.75   & 20.60               & 20.53             & 8.72              & 19.07          \\
\rowcolor{bandB} \multirow{-2}{*}{products}  & GT                     & 25.01           & 6.75          & 109.88            & 2.42          & 73.87    & OOM    & 86.47  & +12.60   & 20.63               & 18.67             & 8.30              & 18.84          \\
\rowcolor{bandA}                             & Graph.                 & 15.95            & 7.31          & 137.5             & 1.10          & 77.55    & OOM    & 76.20  & -1.35    & 20.60               & 19.87             & 14.98             & 33.52          \\
\rowcolor{bandA} \multirow{-2}{*}{Amazon}    & GT                     & 15.88            & 4.34          & 137.66            & 1.62          & 76.82    & OOM    & 72.97  & -3.85    & 20.63               & 19.57             & 14.65             & 34.11          \\
\rowcolor{bandB}                             & Graph.                 & 4.28             & 1.70          & 2.91              & 0.40          & 86.98    & 90.48  & 91.37  & +4.39    & 20.59               & 3.21              & 1.37              & 2.42           \\
\rowcolor{bandB} \multirow{-2}{*}{genius}    & GT                     & 4.22             & 0.75          & 2.8               & 0.54          & 86.66    & 89.78  & 90.42  & +3.76    & 20.62               & 3.95              & 1.40              & 2.00           \\
\rowcolor{bandA}                             & Graph.                 & 30.91            & 11.91         & 32.3              & 1.72          & 31.55    & 54.64  & 61.97  & +30.42   & 20.60               & 20.02             & 5.56              & 6.45           \\
\rowcolor{bandA} \multirow{-2}{*}{snap-pat.} & GT                     & 30.83            & 6.03          & 30.98             & 2.68          & 31.57    & 56.55  & 61.70  & +30.13   & 20.64               & 20.83             & 6.43              & 6.45           \\ \hline
\end{tabular}
\end{table*}

\paragraph{Training efficiency}
  We optimize \torchgt by caching its partitions and sequences across epochs, eliminating repeated construction from its measured training time.
  Even with this optimization, \ours trains $2.2$--$5.6\times$ faster per epoch on every dataset and backbone (\emph{Training} columns of Table~\ref{tab:distributed}).
  We attribute the gap to how the two systems compute attention. 
  Whenever \torchgt's connectivity conditions fail, Dual-interleaved Attention falls back to dense attention, which is quadratic to the sequence length and ignores graph sparsity.
  Its per-epoch time therefore barely changes across backbones ($24.99$ vs.\ $25.01$\,s on \emph{ogbn-products}), whereas \ours bounds attention at $B$ edges per query, so its time reflects the backbone's compute pattern, and GT's lighter layers run $1.5$--$2.3\times$ faster.

  \paragraph{Inference efficiency} 
  \ours also infers $5$--$9\times$ faster on \emph{ogbn-arxiv} and \emph{genius}, with speedups exceeding two orders of magnitude in the best case on the three large graphs (\emph{Inference} columns of Table~\ref{tab:distributed}).
  Most of this gap comes from \torchgt's unoptimized evaluation path, which disables SP, so every rank recomputes all heads over the full node sequence instead of its own partition, and rebuilds each batch's subgraph and fallback condition from scratch.
  \ours instead evaluates through the same sequence-parallel path as training and reuses the cached edge set.
  Since the gap mainly reflects \torchgt's implementation rather than attention, we regard the training speedup as our main efficiency result.
  
  \paragraph{Convergence performance}
  \ours is more accurate on four of the five datasets (Table~\ref{tab:distributed}), by $+13.69$/$+16.75$\,pp on \emph{ogbn-arxiv}, $+10.75$/$+12.60$\,pp on \emph{ogbn-products}, $+4.39$/$+3.76$\,pp on \emph{genius}, and most on \emph{snap-patents}, where it lifts both backbones from about $31.6\%$ to $62\%$.
  The \emph{Full} column separates two sources of this gain.
  Full-graph training contributes most of it, with \emph{Full} beating \torchgt by $+23.09$/$+24.98$\,pp on \emph{snap-patents} and $+14.84$/$+17.19$\,pp on \emph{ogbn-arxiv}, but \emph{Full} runs out of memory on \emph{ogbn-products} and \emph{Amazon}, so the budget is what makes it feasible on large graphs.
  Our mixed scope adds $+7.33$/$+5.15$\,pp over \emph{Full} on \emph{snap-patents} and $+0.89$/$+0.64$\,pp on \emph{genius}, while on homophilic \emph{ogbn-arxiv} keeping all real neighbors is slightly better.
  The exception is \emph{Amazon} ($-1.35$/$-3.85$\,pp), where full-graph training updates the model once per epoch, whereas \torchgt updates it ${\sim}60$ times per epoch in this setting.
  With the same setting extended to 6000 epochs, still with about $20\times$ fewer updates, \ours reaches $77.10\%$/$76.81\%$, matching \torchgt's $77.55\%$/$76.82\%$.
  We therefore attribute the gap to the larger number of updates that a large graph needs, rather than to the budgeted scope.

  \paragraph{Memory usage}
  \ours's GPU memory follows the graph and the budget, whereas \torchgt's stays fixed (\emph{Allocated GPU} and \emph{Peak CPU} columns of Table~\ref{tab:distributed}).
  \torchgt uses almost the same GPU memory on every dataset ($20.59$--$20.64$\,GiB), because once its connectivity conditions fail, memory is set by the sequence length rather than the graph.
  \ours allocates only what the per-query budget needs, $1.85$--$3.95$\,GiB on the small \emph{ogbn-arxiv} and \emph{genius} graphs and $18.67$--$20.83$\,GiB on the three large ones, on par with \torchgt.
  In exchange, \ours caches the sampled edges in host memory, peaking at $34.11$\,GiB with GT on \emph{Amazon} against \torchgt's $14.65$\,GiB, though with GT on \emph{snap-patents} the two nearly match ($6.45$ vs.\ $6.43$\,GiB).
  We consider this a favorable trade, since host memory is far cheaper than HBM and can alleviate the construction bottleneck.

\subsection{Generality on Various Graph Transformer Backbones}
\begin{table}[htbp]
\caption{Native baselines vs.\ their \ours counterparts under budget $B$, all with full activation recomputation.
GPS-F / GPS-P denote GraphGPS with full / Performer attention and Exp native Exphormer. Ours-G / Ours-E replace the attention edges of GraphGPS / Exphormer with our scope. Perf.\ is test accuracy (\%), or ROC-AUC (\%), mean over five seeds, with bold marking the best mean in each group and underlining a tie with it. GPU is peak allocated memory (GiB). OOM marks configurations that exhaust the device.}
\label{tab:generality}
\begin{tabular}{lccccccc}
\hline
\multirow{2}{*}{Model} & \multirow{2}{*}{$B$} & \multicolumn{2}{c}{arxiv} & \multicolumn{2}{c}{genius} & \multicolumn{2}{c}{Amazon} \\ \cline{3-8}
                       &                      & Perf.   & GPU    & Perf.     & GPU    & Perf.   & GPU    \\ \hline
GPS-F                  & --                   & \multicolumn{2}{c}{OOM}       & \multicolumn{2}{c}{OOM}   & \multicolumn{2}{c}{OOM}    \\
GPS-P                  & --                   & 55.89  & 3.57   & 86.72    & 7.06   & \multicolumn{2}{c}{OOM}    \\
Ours-G                 & 8                    & 70.71  & 2.30   & \underline{88.36} & 3.15   & \textbf{75.09} & 22.44  \\
Ours-G                 & 16                   & \textbf{71.36} & 3.16   & \textbf{88.41} & 3.26   & \underline{75.06} & 30.48  \\ \hline
Exp                    & --                   & \textbf{71.67} & 12.21  & \underline{90.10} & 13.97  & \multicolumn{2}{c}{OOM}    \\
Ours-E                 & 8                    & 71.11  & 4.41   & \textbf{90.21} & 6.24   & 73.10  & 40.36  \\
Ours-E                 & 16                   & \underline{71.66} & 6.05   & \underline{90.08} & 6.65   & \multicolumn{2}{c}{OOM}    \\ \hline
\end{tabular}
\end{table}

Our scope carries over to other backbones, matching or beating their native attention with less memory.
Because the controller is model-agnostic, we replace the native attention edges in GraphGPS (full attention) and Exphormer (real + expander edges) with the two edge types it produces, leaving every other component unchanged.
We follow Exphormer's hyperparameters~\cite{exphormer}\footnote{\url{https://github.com/hamed1375/Exphormer/blob/main/configs/Exphormer/arxiv.yaml}} and remove the GNN layers to isolate the effect of the attention scope from local message passing.
All runs, the native ones included, use full activation recomputation, so the memory gaps below come from the attention scope alone.

\paragraph{Versus Performer attention}
Wrapping GraphGPS with our scope improves accuracy and memory at the same time.
The native dense variants are memory-bound (Table~\ref{tab:generality}), as GPS-F OOMs on all three datasets and GPS-P fits only \emph{ogbn-arxiv} and \emph{genius}.
At $B{=}16$, Ours-G leads GPS-P by $15.47$\,pp on \emph{ogbn-arxiv} ($71.36$ vs.\ $55.89$) and by $1.69$\,pp on \emph{genius} ($88.41$ vs.\ $86.72$), at $1.1\times$ and $2.2\times$ less memory.
At $B{=}8$ the memory gap grows to $1.6\times$ on \emph{ogbn-arxiv} while the lead stays at $14.82$ and $1.64$\,pp.
We attribute the gap to candidate choice. Performer approximates attention without using the graph, whereas the controller picks the edges the topology suggests.
Ours-G also trains on \emph{Amazon} ($75.09$), where GPS-P does not run at all.

\paragraph{Versus Exphormer}
Against Exphormer, a far stronger but costlier baseline, our scope matches its accuracy with about half the memory.
Native Exphormer needs $12.21$ and $13.97$\,GiB on \emph{ogbn-arxiv} and \emph{genius} and OOMs on \emph{Amazon}.
At $B{=}16$, Ours-E matches it on both graphs ($71.66$ vs.\ $71.67$ and $90.08$ vs.\ $90.10$) with $2.0\times$ and $2.1\times$ less memory.
At $B{=}8$ it even posts the best mean on \emph{genius} ($90.21$) with $2.2\times$ less memory.
Without its expander edges, native Exphormer drops to $87.78$ on \emph{genius}.
Ours-E has none and matches the full Exphormer with random-walk edges.
Ours-E also trains on \emph{Amazon} at $B{=}8$ ($73.10$).

\paragraph{Where the memory goes}
\ours materializes at most $B$ edges per query instead of the whole neighborhood, so its attention never grows with the graph's dense regions.
This is why Ours-E uses less memory than Exphormer even though Exphormer's scope is already sparse.
When a budget does not fit on one GPU, more devices make it fit without lowering $B$. Ours-E at $B{=}16$ OOMs on \emph{Amazon} on one GPU but trains on two at $\mathrm{SP}{=}2$ ($73.19\%$), an option neither native baseline offers.

\subsection{Distributed Scalability and Capacity}
\label{sec:exp_scalability}

We scale \ours with the number of devices, the graph, and the budget $B$, using the even split $b_r{=}b_w{=}B/2$, $L{=}4$, and $R{=}2$.
All runs use full recomputation, so every configuration uses the same activation policy and the results are directly comparable.
We report the mean epoch time after warm-up and the maximum per-rank peak memory.

\begin{figure}[t]
\centering
\begin{subfigure}[t]{\linewidth}
  \centering
  \includegraphics[width=\linewidth]{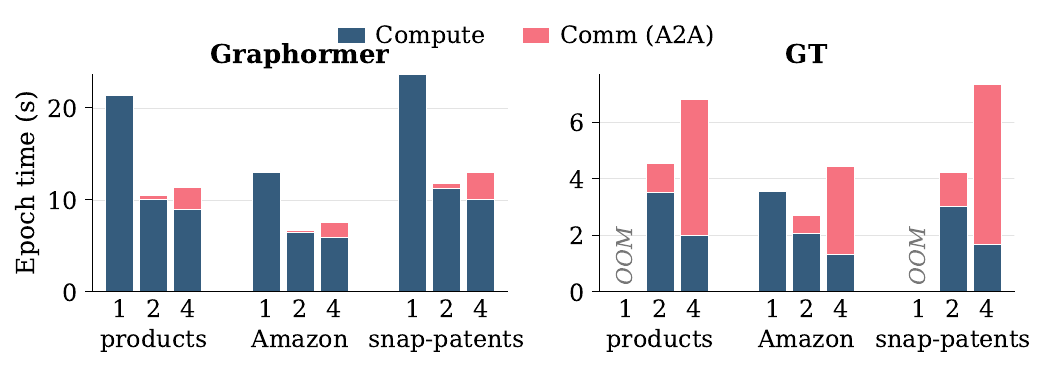}
  \caption{Epoch time split by compute and all-to-all communication.}
  \label{fig:sp_time}
\end{subfigure}
\vspace{0.3em}
\begin{subfigure}[t]{\linewidth}
  \centering
  \includegraphics[width=\textwidth]{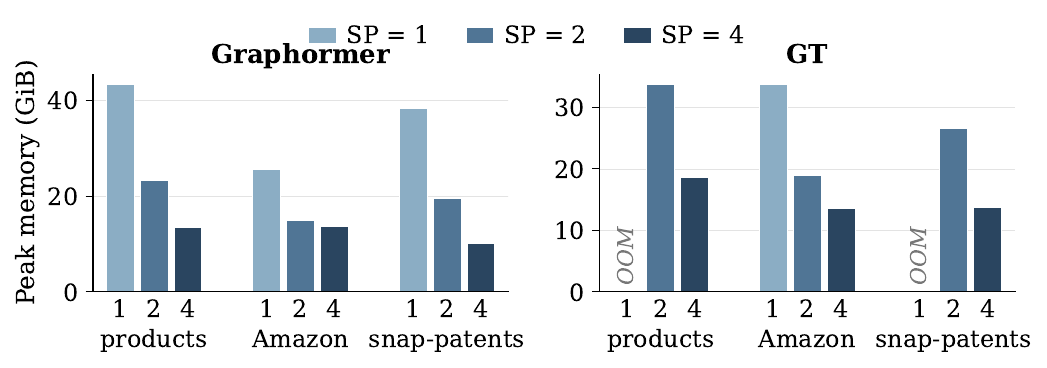}
  \caption{Per-rank peak GPU memory allocation.}
  \label{fig:sp_memory}
\end{subfigure}
\caption{Strong scaling at $B{=}8$ with $\mathrm{SP}\in\{1,2,4\}$. Within each dataset group, bars correspond to SP degrees 1, 2, and~4. OOM marks out-of-memory cases on GPU.}
\label{fig:sp_scaling}
\end{figure}

\paragraph{Scaling with devices}
Adding devices shortens training up to $\mathrm{SP}{=}2$. Beyond it, all-to-all communication outweighs the speedup, although per-rank memory keeps falling.
Fig.~\ref{fig:sp_scaling} fixes $B{=}8$ and varies $\mathrm{SP}\in\{1,2,4\}$.
For Graphormer, going from $\mathrm{SP}{=}1$ to $2$ cuts epoch time by $1.94$--$2.04\times$ on the three graphs, and all-to-all communication takes only $4$--$5\%$ of it.
Going to $\mathrm{SP}{=}4$ raises the communication share to $21$--$23\%$, so the epoch is $9$--$12\%$ slower than at $\mathrm{SP}{=}2$, although per-rank memory keeps falling, by $3.2\times$ on \emph{ogbn-products} and $3.8\times$ on \emph{snap-patents} from $\mathrm{SP}{=}1$ to $4$ (Fig.~\ref{fig:sp_scaling}(b)).
GT runs out of memory at $\mathrm{SP}{=}1$ on two graphs.
Its layers are lighter, so communication weighs more. At $\mathrm{SP}{=}4$, all-to-all takes $70$--$77\%$ of the epoch, making $\mathrm{SP}{=}4$ slower than $\mathrm{SP}{=}2$ even though it cuts per-rank memory by another $1.4$--$2.0\times$.

\paragraph{Cost bound}
Since each node reads at most $B$ attention edges, the cost of an epoch should grow with $nB$, not with the number of edges in the graph.
Both factors show up in our measurements.
At $B{=}8$, epoch time follows the number of nodes rather than edges (Fig.~\ref{fig:sp_scaling}(a)).
With $19\times$ fewer edges than \emph{Amazon} but $1.9\times$ more nodes, \emph{snap-patents} takes $1.7$--$1.8\times$ longer per epoch for Graphormer and $1.6$--$1.7\times$ for GT.
On a fixed graph and SP degree, epoch time grows at most linearly with $B$. On \emph{Amazon} at $\mathrm{SP}{=}4$, raising $B$ from 8 to 64 lengthens the epoch by only $15\%$ for Graphormer ($7.51{\to}8.66$\,s) and $40\%$ for GT ($4.41{\to}6.18$\,s), because attention is only part of each layer's work and many queries have fewer than $B$ distinct candidates.
In contrast, \torchgt's dense fallback grows quadratically with the sequence length.

\paragraph{Capacity under a memory budget}
With a 24\,GiB/rank memory budget on \emph{Amazon}/Graphormer, the largest $B$ that fits is $8$ at $\mathrm{SP}{=}1$, $64$ at $\mathrm{SP}{=}2$, and at least $128$ at $\mathrm{SP}{=}4$.
This capacity is predictable.
In the ten runs that fit, per-rank peak memory is linear in the number of attention edges actually built and inversely proportional to the SP degree, with a fitting error of at most $1.5\%$.
Since many nodes have fewer than $B$ distinct candidates, this number grows more slowly than $nB$.
At $\mathrm{SP}{=}4$, raising $B$ from $8$ to $128$ increases peak memory by only $2.9\times$ ($5.5{\to}15.8$\,GiB).
The fit also predicts that the two runs that fail, $B{=}16$ at $\mathrm{SP}{=}1$ and $B{=}128$ at $\mathrm{SP}{=}2$, exceed the memory budget.

\subsection{Effectiveness of Logical Planning}
\label{sec:exp_adaptive}

\begin{figure}[t]
  \centering
  \includegraphics[width=\columnwidth]{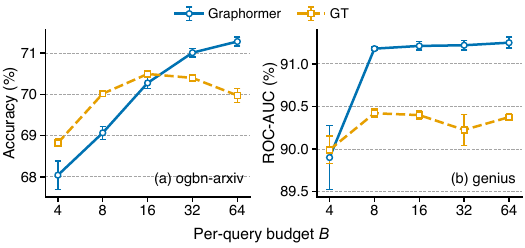}
  \caption{Accuracy (\%) on \emph{ogbn-arxiv} and ROC-AUC (\%) on \emph{genius} as the per-query budget $B$ grows, with \emph{Adaptive} ($L{=}4$, $R{=}2$). Means and standard errors are over five seeds.}
  \label{fig:budget_curve}
\end{figure}

We evaluate attention-scope effectiveness, the controller's allocation quality and overhead, and sensitivity to budget size, walk parameters, and controller settings.
We study \emph{ogbn-arxiv} and \emph{genius}, whose accuracy is sensitive to the real/RW split (Table~\ref{tab:adaptive}).
Tables~\ref{tab:scope_comparison} and~\ref{tab:adaptive} and Fig.~\ref{fig:budget_curve} report means over five seeds.
\emph{Tied} means a gap within twice its standard error, i.e., within seed-to-seed noise.

\paragraph{Comparison with other scopes}
Our adaptive scope is best or tied in seven of the eight backbone--dataset pairs of Table~\ref{tab:scope_comparison} without per-dataset tuning, whereas no fixed scope is best on both graphs.
At $B{=}16$, \emph{Adaptive} is our mixed scope, whose real/RW split the controller selects.
\emph{All real} and \emph{All RW} draw every candidate from one pool, and \emph{$k$-hop} takes up to $f$ neighbors per hop over $k$ hops, i.e., $\sum_{i=1}^{k}f^i>16$ candidates.
\begin{table}[t]
\centering
\setlength{\tabcolsep}{4pt}
\caption{Test accuracy (\%) on \emph{ogbn-arxiv} and ROC-AUC (\%) on \emph{genius} under different attention scopes across four backbones, mean over five seeds.
Bold marks the best scope for each backbone and dataset, and underlining a scope tied with it.}
\label{tab:scope_comparison}
\begin{tabular}{cccccc}
\hline
Dataset                 & Scope           & Graph.          & GT                 & GraphGPS           & Exphormer          \\ \hline
\multirow{5}{*}{arxiv}  & All real        & \textbf{71.43}  & \textbf{70.59}     & \textbf{71.32}     & \underline{71.60}  \\
                        & All RW          & 66.42           & 66.30              & 66.81              & 66.87              \\
                        & 2-hop $(f{=}4)$ & 70.08           & 69.51              & 70.09              & 70.54              \\
                        & 3-hop $(f{=}3)$ & 69.37           & 68.96              & 69.83              & 69.67              \\
                        & Adaptive        & 70.28           & \underline{70.49}  & \underline{71.22}  & \textbf{71.69}     \\ \hline
\multirow{5}{*}{genius} & All real        & 90.30           & 89.50              & \underline{88.43}  & 88.67              \\
                        & All RW          & 89.37           & 88.55              & 87.96              & 88.90              \\
                        & 2-hop $(f{=}4)$ & 89.54           & 88.85              & 87.82              & 89.33              \\
                        & 3-hop $(f{=}3)$ & 89.43           & 88.77              & 87.75              & 89.15              \\
                        & Adaptive        & \textbf{91.21}  & \textbf{90.40}     & \textbf{88.49}     & \textbf{90.32}     \\ \hline
\end{tabular}
\end{table}

On homophilic \emph{ogbn-arxiv}, \emph{Adaptive} matches \emph{All real} on GT, GraphGPS, and Exphormer and falls $1.15$\,pp behind it on Graphormer.
On heterophilic \emph{genius}, it leads \emph{All real} by $0.90$--$1.65$\,pp on three backbones and matches it on GraphGPS.
\emph{All RW} and both \emph{$k$-hop} scopes fall behind \emph{Adaptive} on every backbone, so GNN-style $k$-hop expansion transfers poorly to Graph Transformer instances.

  \begin{table}[t]
  \centering
  \caption{Allocation quality at $B{=}8$ ($L{=}4$, $R{=}2$): accuracy (\%) or ROC-AUC (\%) for each split is the mean of five seeds.
  Bold: best in each row; underlined: tied with it.
  Superscript: number of seeds in which the controller froze at that split.}
  \label{tab:adaptive}
\setlength{\tabcolsep}{3pt}
\begin{tabular}{@{}clrrrrr}
\hline
\multirow{2}{*}{Dataset} & \multirow{2}{*}{Model} & \multicolumn{5}{c}{Fixed $(b_r,b_w)$} \\ \cline{3-7}
 & & $(0,8)$ & $(2,6)$ & $(4,4)$ & $(6,2)$ & $(8,0)$ \\ \hline
\multirow{4}{*}{arxiv} & Graph.    & 66.54    & 67.73                            & 69.09\rlap{$^{5}$}               & 70.19                            & \textbf{70.86} \\
                       & GT        & 66.29    & 67.79                            & 68.88                            & 69.69                            & \textbf{70.07}\rlap{$^{5}$} \\
                       & GraphGPS  & 66.75    & 68.12                            & 69.14                            & 70.00                            & \textbf{70.69}\rlap{$^{5}$} \\
                       & Exphormer & 66.89    & 68.94                            & 70.23                            & 70.72                            & \textbf{71.05}\rlap{$^{5}$} \\ \hline
\multirow{4}{*}{genius} & Graph.    & 89.35    & \textbf{91.23}\rlap{$^{2}$}      & \textbf{91.23}\rlap{$^{1}$}      & \underline{91.20}\rlap{$^{2}$}   & 90.30 \\
                       & GT        & 88.53    & \textbf{90.29}                   & \underline{90.25}\rlap{$^{5}$}   & \underline{90.27}                & 89.35 \\
                       & GraphGPS  & 87.97    & \underline{88.36}                & \underline{88.35}\rlap{$^{5}$}   & \textbf{88.42}                   & \underline{88.40} \\
                       & Exphormer & 88.91    & \underline{90.18}\rlap{$^{1}$}   & \underline{90.23}\rlap{$^{4}$}   & \textbf{90.32}                   & 88.76 \\ \hline
\end{tabular}
\end{table}
\paragraph{Allocation quality}
The controller freezes at the best split, or one tied with it, in almost all settings.
Table~\ref{tab:adaptive} compares the split at which the controller freezes with five fixed splits at $B{=}8$.
On \emph{ogbn-arxiv} the best split, $(8,0)$, is unique, and the controller picks it in every seed for GT, GraphGPS, and Exphormer.
On \emph{genius}, the three mixed splits are tied (within $0.14$\,pp) while the extreme ones lose up to $1.88$\,pp, and the controller picks a mixed split in all $20$ runs.
Its trained models stay within $0.05$\,pp of the best split on \emph{ogbn-arxiv} and $0.31$\,pp on \emph{genius}, except for Graphormer on \emph{ogbn-arxiv}, which freezes at $(4,4)$ and loses $1.77$\,pp at $B{=}8$ and $1.15$\,pp at $B{=}16$.
We believe the cause is its global virtual node, which only Graphormer has. 
Early in training, it gives every query a graph-wide summary, so the choice of local keys barely changes the probe loss, and the controller freezes before the informative real neighbors are aggregated.

\begin{figure}[t]
  \centering
  \includegraphics[width=\columnwidth]{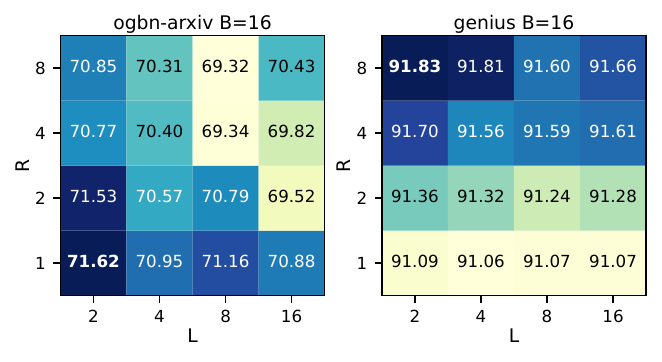}
  \caption{Graphormer sensitivity over the $L\times R$ grid at $B{=}16$.
  Values are test accuracy (\%) for \emph{ogbn-arxiv} and
  ROC-AUC (\%) for \emph{genius} from a single seed. Bold marks the best.}
  \label{fig:heatmap_b16}
\end{figure}

\paragraph{Walk-parameter sensitivity}\looseness=-1
Using fewer walks per query works better on \emph{ogbn-arxiv}, whereas \emph{genius} benefits from more walks.
We sweep the walk length $L$ over $\{2,4,8,16\}$ and the number of walks $R$ over $\{1,2,4,8\}$ at $B{=}16$ with Graphormer (Fig.~\ref{fig:heatmap_b16}).
Both datasets respond to $R$, as \emph{ogbn-arxiv} peaks at $(L,R){=}(2,1)$ ($71.62\%$) and degrades as $R$ grows, whereas \emph{genius} peaks at $(2,8)$ ($91.83$) and improves with it.
This is consistent with the long-range attention in Fig.~\ref{fig:hop_relative}(a). On \emph{ogbn-arxiv}, attention beyond the first hop falls far below the uniform level, so extra walks reach nodes the model barely attends to, whereas \emph{genius} keeps benefiting from additional candidates.

\paragraph{Budget size}
A small budget is usually enough.
Fig.~\ref{fig:budget_curve} varies $B$ from $4$ to $64$ for Graphormer and GT.
Raising $B$ from $4$ to $8$ helps in every case.
Beyond that, the trend depends on the backbone rather than the dataset.
Graphormer never loses accuracy as $B$ grows, whereas GT peaks at a moderate budget ($B{=}16$ on \emph{ogbn-arxiv}, $B{=}8$ on \emph{genius}) and then declines slightly.
Results also become less sensitive to the seed, as the standard error drops from $0.09$--$0.37$\,pp at $B{=}4$ to $0.02$--$0.16$\,pp at $B{=}8$ and stays there, except for GT at $B{=}32$ on \emph{genius}.
In three of the four cases, $B{\le}16$ already reaches or matches the best, so the memory it costs, can stay low.

\paragraph{Selection overhead}
The controller needs only one training run. Across the runs in Table~\ref{tab:distributed}, it freezes the split within about $30$ epochs, and probing costs $0.3\%$--$9.7\%$ of the 2000-epoch schedule (median $1.5\%$).
Exhaustive search would need five full runs at $B{=}8$, one per split.
The controller selects a near-best split in almost all settings (Table~\ref{tab:adaptive}).

\paragraph{Controller sensitivity}
The controller needs no threshold tuning.
Varying $\tau\in\{10^{-5},10^{-4}\}$, $K_{\max}\in\{2,4\}$, or $P\in\{2,5\}$ one at a time gives six settings.
Over Graphormer and GT, both graphs, and three seeds at $B{=}8$ ($72$ runs), every mean is tied with the default's, and $61$ runs freeze at the same split as the default.
Apart from the Graphormer/\emph{ogbn-arxiv} exception above, $98\%$ of runs ($53$ of $54$) freeze at a split tied with the best fixed split, showing that the controller is stable.

\subsection{Effectiveness of Physical Planning}
\label{sec:exp_planner}

We evaluate training feasibility and epoch time against two fixed baselines, \emph{gpurec} (GPU build, full recomputation) and \emph{cpu-pf} (CPU build with prefetch), on Graphormer and GT, the most and least compute-intensive of our four backbones.
The grid spans three HBM budgets ($16$/$24$/$40$\,GiB), two depths ($D{=}4,8$), and two datasets, at $B{=}8$, $\mathrm{SP}{=}4$.

\begin{table*}[!t]
\begin{minipage}[t]{0.485\textwidth}
\centering
\caption{Per-epoch time (s) at a $16$\,GiB budget ($B{=}8$, $\mathrm{SP}{=}4$,
$(b_r,b_w){=}(4,4)$). OOM marks peak${>}$budget.
Path is the edge-construction path the planner commits.}
\label{tab:planner_feasibility}
\setlength{\tabcolsep}{4pt}
\begin{tabular}{@{}llc ccc l@{}}
\toprule
& & & \multicolumn{3}{c}{Epoch (s)} & \\
\cmidrule(lr){4-6}
Dataset & Model & $D$ & gpurec & cpu-pf & planner & Path \\
\midrule
\multirow{4}{*}{Amazon} & \multirow{2}{*}{Graph.}
   & 4 & 7.5  & 11.9 & \textbf{7.2}  & GPU-pers. \\
 & & 8 & 14.1 & 13.3 & \textbf{13.0} & CPU-bcast \\
 & \multirow{2}{*}{GT}
   & 4 & 4.4  & 11.5 & \textbf{4.0}  & GPU-pers. \\
 & & 8 & 7.8  & 11.5 & \textbf{7.5}  & GPU-pers. \\
\midrule
\multirow{4}{*}{snap-pat.} & \multirow{2}{*}{Graph.}
   & 4 & 13.0 & 12.5 & \textbf{12.2} & CPU-local \\
 & & 8 & 25.2 & 24.7 & \textbf{24.4} & CPU-local \\
 & \multirow{2}{*}{GT}
   & 4 & OOM  & 7.1  & \textbf{6.5}  & CPU-local \\
 & & 8 & OOM  & 14.0 & \textbf{13.9} & CPU-local \\
\bottomrule
\end{tabular}
\end{minipage}\hfill
\begin{minipage}[t]{0.485\textwidth}
\centering
\caption{Planner retention plans in Table~\ref{tab:planner_feasibility}'s setting.
Tiers count recomputed, attention-retaining, and retained layers,
$D_{\mathrm{rec}}/D_{\mathrm{mha}}/D_{\mathrm{ret}}$. Budgets are in GiB.}
\label{tab:planner_retention}
\setlength{\tabcolsep}{4pt}
\begin{tabular}{@{}llc ccc ccc@{}}
\toprule
& & & \multicolumn{3}{c}{Tiers} & \multicolumn{3}{c}{Epoch (s)} \\
\cmidrule(lr){4-6} \cmidrule(lr){7-9}
Dataset & Model & $D$ & 16 & 24 & 40 & 16 & 24 & 40 \\
\midrule
\multirow{4}{*}{Amazon} & \multirow{2}{*}{Graph.}
   & 4 & 2/0/2 & 0/0/4 & 0/0/4 & 7.2  & 7.0  & 7.0  \\
 & & 8 & 5/0/3 & 2/4/2 & 0/1/7 & 13.0 & 12.8 & 12.6 \\
 & \multirow{2}{*}{GT}
   & 4 & 3/0/1 & 1/1/2 & 0/0/4 & 4.0  & 3.6  & 3.5  \\
 & & 8 & 7/0/1 & 5/1/2 & 2/3/3 & 7.5  & 7.2  & 6.6  \\
\midrule
\multirow{4}{*}{snap-pat.} & \multirow{2}{*}{Graph.}
   & 4 & 3/0/1 & 1/1/2 & 0/0/4 & 12.2 & 11.8 & 11.7 \\
 & & 8 & 7/0/1 & 6/0/2 & 4/0/4 & 24.4 & 24.1 & 23.9 \\
 & \multirow{2}{*}{GT}
   & 4 & 3/0/1 & 2/1/1 & 0/3/1 & 6.5  & 5.9  & 5.2  \\
 & & 8 & 8/0/0 & 7/0/1 & 5/0/3 & 13.9 & 12.6 & 12.0 \\
\bottomrule
\end{tabular}
\end{minipage}
\end{table*}

\paragraph{Fixed strategies versus the planner}
Across all configurations, the planner selects feasible plans and achieves lower epoch times than every feasible strategy.
Table~\ref{tab:planner_feasibility} reports the $16$\,GiB results.
Here, \emph{gpurec} exceeds the budget for GT on \emph{snap-patents} at both depths and is slower than the planner elsewhere.
In contrast, \emph{cpu-pf} always fits but exposes construction latency when the computation window is short.
The planner meets the budget through phase-aware memory estimation (Eq.~\eqref{eq:phase_memory}) and edge-path selection.
For GT on \emph{Amazon} at $D{=}4$, it takes $4.0$\,s per epoch versus $11.5$\,s for \emph{cpu-pf} ($2.9\times$ speedup) and $4.4$\,s for \emph{gpurec}.
The preferred strategy also depends on the split. 
Additional runs on \emph{Amazon} at $D{=}8$ and $16$\,GiB show that GT's \emph{gpurec} takes $0.68\times$ the epoch time of \emph{cpu-pf} at $(4,4)$ but $1.04\times$ at $(0,8)$.
The planner is fastest across $(0,8)$, $(4,4)$, and $(8,0)$ ($0.65$--$0.98\times$ \emph{cpu-pf}'s time), supporting workload-aware planning rather than selection from $B$ alone.

\begin{figure}[!t]
\centering
\includegraphics[width=\linewidth]{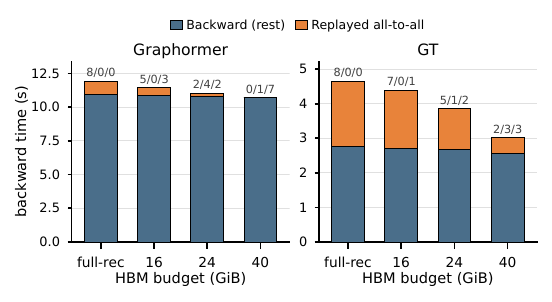}
\caption{Backward time as the HBM budget grows (\emph{Amazon}, $D{=}8$, $B{=}8$, $\mathrm{SP}{=}4$). Orange marks the all-to-all calls replayed.
Labels give the committed $D_{\mathrm{rec}}/D_{\mathrm{mha}}/D_{\mathrm{ret}}$.}
\label{fig:planner_comm}
\end{figure}

\paragraph{Benefit of extra HBM}
Increasing the HBM budget reduces epoch time by up to $20\%$ for GT ($6.5{\to}5.2$\,s on \emph{snap-patents} at $D{=}4$) and $4\%$ for Graphormer (Table~\ref{tab:planner_retention}).
The planner retains more states and recomputes fewer layers.
For GT on \emph{Amazon} at $D{=}8$, the retention layout changes from $7/0/1$ to $2/3/3$.
Under SP, this also reduces communication, because each layer issues four all-to-all calls per forward pass and recomputation raises the total from $4D$ to $4(D{+}D_{\mathrm{rec}})$ (Fig.~\ref{fig:planner_comm}).
Replayed calls account for $41\%$ of GT's backward time but only $8\%$ of Graphormer's, whose layers are more compute-intensive.
Moving from full recomputation to the $40$\,GiB plan reduces backward time by $35\%$ for GT and $10\%$ for Graphormer.
Backward time grows linearly with $D_{\mathrm{rec}}$ (residuals below $1\%$), allowing a few calibration layers to estimate every retention layout.
Additional devices also provide memory headroom. In Fig.~\ref{fig:sp_scaling}'s strong-scaling setting at $\mathrm{SP}{=}4$, the planner with a $24$\,GiB/rank budget reduces epoch time by $5$--$7\%$ for Graphormer and $7$--$16\%$ for GT relative to full recomputation.
Rerunning the planner at $10$\,GiB on \emph{Amazon} shows what its cost model captures.
Compared with $40$\,GiB, Graphormer at $D{=}8$ allocates $8.5$ rather than $28.9$\,GiB and slows by only $3\%$, since its $13$\,s computation still hides the $11$\,s CPU construction (Fig.~\ref{fig:cpu_prefetch_overlap}).
GT at $D{=}4$ must move construction from the GPU to the CPU, as the $13.6$\,GiB GPU peak no longer fits.
Its $3$\,s computation hides little, so epochs slow $3.3\times$, but training still completes.
\paragraph{Plan choice and profiling cost}
No edge path is fastest across all workloads (\emph{Path} in Table~\ref{tab:planner_feasibility}), so the planner profiles each path and selects the fastest feasible one.
On \emph{snap-patents}, CPU construction takes $2$--$3$\,s and is fully hidden by computation, favoring \emph{CPU rank-local prefetch}.
On \emph{Amazon}, CPU construction is slower and GPU construction usually wins, with two exceptions.
Graphormer at $D{=}8$ computes long enough to hide \emph{CPU broadcast prefetch}, and the additional $(0,8)$ runs, whose RW-only split makes construction cheap, use \emph{CPU rank-local prefetch} on both backbones.
The \emph{GPU ephemeral} path is selected when persistent topology exceeds the budget but GPU construction remains fastest (e.g., \emph{Amazon}/Graphormer at $B{=}128$ in Section~\ref{sec:exp_scalability}).
Profiling is a one-time and negligible cost.
It occupies a single epoch, which builds edges once per path and adds $1$--$86$\,s, only $0.07\%$--$0.89\%$ of a 2000-epoch run in Table~\ref{tab:distributed}.
All other epochs train normally, using the selected plan once committed.
\section{Conclusion}
We presented \ours, a system for distributed full-graph training of Graph Transformers that separates a logical plan of which attention edges each query materializes from a physical plan of how to build them within a memory budget, two concerns that prior work conflates.
A probe-guided controller allocates each query's attention budget between real graph edges and random-walk edges and freezes the split once it stabilizes, without per-dataset tuning.
A cost-based planner then compiles the stabilized logical plan into a physical plan.
It assigns edge construction to CPU or GPU, overlaps construction with computation, and adjusts activation retention to fit within memory.
Experiments on five graphs with four backbones confirm that \ours reaches comparable or up to $30.4$ percentage points higher accuracy than \torchgt, trains up to $5.6\times$ faster per epoch, and reduces peak GPU memory by up to an order of magnitude.
It also trains GraphGPS and Exphormer on graphs where their native implementations run out of memory.
Future work will adapt the walk configuration during training.

\section*{Acknowledgment}

We used ChatGPT to assist with refining code logic and data processing for the experiments section. 
All AI-assisted code and processing outputs were reviewed and verified by the authors.

\bibliographystyle{IEEEtran}
\bibliography{refs}

\end{document}